# Interlayer Coupling-Induced Quantum Phase Transition in Quantum Anomalous Hall Multilayers

Ling-Jie Zhou[1,2], Deyi Zhuo[1,2], Ruobing Mei[1], Yi-Fan Zhao[1], Kaijie Yang[1], Ruoxi Zhang[1], Zijie Yan[1], Han Tay[1], Moses H. W. Chan[1], Chao-Xing Liu[1], and Cui-Zu Chang[1]

[1] Department of Physics, The Pennsylvania State University, University Park, PA 16802, USA

[2] These authors contributed equally: Ling-Jie Zhou and Deyi Zhuo

Corresponding authors: cxc955@psu.edu (C.-Z. C.)

**Abstract: A quantum phase transition arises from competition between different ground states and is typically accessed by varying a single physical parameter near absolute zero temperature. The quantum anomalous Hall (QAH) effect with high Chern number $C$ has recently been achieved in magnetic topological insulator (TI) multilayers. In this work, we employ molecular beam epitaxy to synthesize a series of magnetic TI penta-layers by varying the thickness of the middle magnetic TI layer, designated as $m$ quintuple layers. Electrical transport measurements demonstrate a quantum phase transition between $C = 1$ and $C = 2$ QAH states. For $m \leq 1$ and $m \geq 2$, the sample exhibits the well-quantized $C = 1$ and $C = 2$ QAH states, respectively. For $1 < m < 2$, we observe a monotonic decrease in Hall resistance from $h/e^2$ to $h/2e^2$ with increasing $m$, accompanied by a peak in the longitudinal resistance. The quantum phase transition between $C = 1$ and $C = 2$ QAH states is attributed to the weakening of the interlayer coupling between the top and the bottom $C = 1$ QAH layers. Our findings provide a scalable strategy for engineering QAH devices with a tunable Chern number. This approach enables precise control and enhanced functionality in chiral edge current-based electronic devices.**

**Main text:** The exploration of quantum phase transitions and emergent topological states is a vibrant field in condensed matter physics [1-3]. A prime example is the plateau-to-plateau transition in the quantum Hall (QH) effect, where the filling factor of each plateau can be controlled by sweeping an external magnetic field or by adjusting the carrier density of the sample [4-6]. The quantum anomalous Hall (QAH) effect is the zero magnetic field counterpart of the integer QH effect. Like the QH effect, the QAH effect harbors dissipationless chiral edge states with quantized Hall resistance $h/Ce^2$ and vanishing longitudinal resistance at zero magnetic field [7-12]. Here, $C$ is the Chern number of the QAH state, corresponding to the number of chiral edge states. The QAH effect has been realized in magnetically doped TI, including Cr- and/or V-doped $(Bi,Sb)_2Te_3$ thin films [11-16], intrinsic magnetic TI $MnBi_2Te_4$ [17], and moiré materials [18-21]. Owing to its resistance-free chiral edge channels, the QAH insulator is a promising platform for energy-efficient electronics, spintronics, and topological quantum computations [7].

The QAH effect with $C$ of 1 to 5 has been recently realized in magnetic TI multilayers, which are constructed by sequentially stacking 3 QL heavily Cr-doped $(Bi,Sb)_2Te_3$ and 4 QL $(Bi,Sb)_2Te_3$ using molecular beam epitaxy (MBE) [22]. The heavily Cr-doped $(Bi,Sb)_2Te_3$ layer plays a two-fold role in these QAH multilayers: (*i*) the heavy Cr doping concentration reduces the spin-orbit coupling, transforming the Cr-doped $(Bi,Sb)_2Te_3$ layer into a trivial ferromagnetic insulator. (*ii*) it breaks the time-reversal symmetry of the undoped $(Bi,Sb)_2Te_3$ layer, thereby creating the $C$=1 QAH effect. The Chern number $C$ can be tuned by adjusting the thickness of the middle magnetic TI layer [22]. Our prior work [22] has focused on the integer $m$ and well-quantized QAH states without examining the plateau phase transition region between $C = 1$ and $C = 2$ QAH states at zero magnetic field. Specifically, the $m$=1 sample exhibits the $C = 1$ QAH state, while the $m = 2$ sample shows the $C = 2$ QAH state. A change of only 1 QL can drive a plateau phase transition between $C = 1$ and $C = 2$ QAH states. The physical mechanism behind this plateau phase transition and how the second chiral edge channel emerges with varying $m$ remain unexplored.

In this work, we employ MBE to grow a series of magnetic TI penta-layers, specifically 3 QL $(Bi,Sb)_{1.73}Cr_{0.27}Te_3$/4 QL $(Bi,Sb)_2Te_3$/$m$QL $(Bi,Sb)_{1.73}Cr_{0.27}Te_3$/4 QL $(Bi,Sb)_2Te_3$/3 QL $(Bi,Sb)_{1.73}Cr_{0.27}Te_3$ with different $m$ (Figs. 1a and S1)[23]. For magnetic TI sandwiches, a thickness of 4 QL has been demonstrated to be optimal for the middle undoped TI layer to realize the well-quantized QAH state [7,26]. This thickness effectively balances the contributions of bulk and/or helical edge states, which become more pronounced in thicker samples while avoiding the formation of a large hybridization gap that emerges in thinner samples. Meanwhile, a minimum thickness of 3 QL is required for the two magnetic doped TI layers to achieve perpendicular magnetic anisotropy, which is crucial for opening magnetic exchange gaps on the top and the bottom surfaces of the middle undoped TI layer.

By systematically varying $m$, a quantum phase transition between $C = 1$ and $C = 2$ QAH states occurs under zero magnetic field. For $m \leq 1$ and $m \geq 2$, the samples exhibit well-quantized $C = 1$ and $C = 2$ QAH states, respectively. For $1 < m < 2$, as $m$ increases, the Hall resistance $\rho_{yx}$ monotonically decreases from $\sim h/e^2$ to $\sim h/2e^2$, while the longitudinal resistance $\rho_{xx}$ shows a peak feature. The critical point of the quantum phase transition is found at $m = 1.7$. Moreover, the longitudinal and Hall resistance ratio at zero magnetic field $\rho_{yx}(0)/\rho_{xx}(0)$ exceeds 1 across the entire plateau phase transition. This observation indicates that during the plateau phase transition, a second chiral edge channel emerges, while the chiral edge channel of the original $C = 1$ QAH state persists [7].

The magnetic TI penta-layers used in this work are grown on heat-treated insulating $SrTiO_3$ (111) substrates in an MBE chamber (Omicron Lab10) with a vacuum better than $\sim 2 \times 10^{-10}$ mbar.

The Cr doping concentration $x$ in the three magnetic TI layers is fixed at $x = 0.27$, situating the magnetic TI layer in the trivial insulator regime [22,27-30]. The Bi/Sb ratio in each layer is optimized to tune the chemical potential of the entire sample near the charge-neutral point. The chemical potential is further finely adjusted using a bottom gate voltage $V_g$. The thickness $m$ of the middle magnetic TI layer is determined by the MBE growth duration. Electrical transport measurements are conducted using both a physical property measurement system (PPMS, Quantum Design DynaCool, 2K, 9T) for $T \geq 1.7$ K and a dilution refrigerator (Leiden Cryogenics, 10mK, 9T) for $T < 1.7$ K. The magnetic field is applied perpendicular to the film plane. All penta-layer samples are scratched into a Hall bar geometry using a computer-controlled probe station. More details about the MBE growth and transport measurements can be found in [23].

We first perform magneto transport measurements on a series of magnetic TI penta-layers with different $m$ at $T = 25$ mK and charge neutral point $V_g = V_g^0$ (Figs. 1b, 1c, and S2) [23]. The value of $m$ varies from 0 to 3 across 14 devices, specifically $m = 0$, 0.4, 0.8, 1, 1.1, 1.2, 1.3, 1.5, 1.6, 1.7, 1.8, 2, 2.5, and 3. The $m$ value is determined by the MBE growth duration and further calibrated through atomic force microscopy measurements. The fractional $m$ value implies $m \times 100\%$ coverage of 1 QL magnetic TI layer [31,32]. The $V_g^0$ value is determined at the voltage where the Hall resistance under zero magnetic field [labeled $\rho_{yx}(0)$] is maximized. For devices with $0 \leq m \leq 1$, the well-quantized $C = 1$ QAH state appears (Fig. 1b). The values of $\rho_{yx}(0)$ are ~1.0041 $h/e^2$, ~1.0105 $h/e^2$, ~0.9951 $h/e^2$, and ~0.9839 $h/e^2$ for $m = 0$, 0.4, 0.8, and 1, respectively. The corresponding longitudinal resistance under zero magnetic field [labeled $\rho_{xx}(0)$] are ~0.0013 $h/e^2$ (~34 Ω), ~0.0001 $h/e^2$ (~2 Ω), ~0.0024 $h/e^2$ (~63 Ω), and ~0.0023 $h/e^2$ (~59 Ω) (Figs. S2a to S2d) [23]. As $m$ increases from 1 to 2, i.e., $1 < m < 2$, the sample deviates from the well-quantized $C = 1$ QAH state and progressively approaches the well-quantized $C = 2$ QAH state (Fig. 1c). $\rho_{yx}(0)$ gradually changes from ~$h/e^2$ to ~$h/2e^2$, while $\rho_{xx}(0)$ first increases and then decreases back to nearly zero. The values of $\rho_{yx}(0)$ are ~0.9803 $h/e^2$, ~0.9706 $h/e^2$, ~0.9491 $h/e^2$, ~0.9195 $h/e^2$, ~0.6542 $h/e^2$, ~0.5687 $h/e^2$, and ~0.5071 $h/e^2$ for $m$ =1.1, 1.2, 1.3, 1.5, 1.6, 1.7, and 1.8, respectively. The corresponding $\rho_{xx}(0)$ are ~0.0150 $h/e^2$, ~0.0589 $h/e^2$, ~0.1311 $h/e^2$, ~0.1522 $h/e^2$, ~0.3125 $h/e^2$, ~0.3745 $h/e^2$, and ~0.2357 $h/e^2$ (Figs. S2e to S2k) [23].

As $m$ further increases from 2 to 3, i.e., $2 \leq m \leq 3$, the samples stabilize at the well-quantized $C = 2$ QAH state (Fig. 1d). The values of $\rho_{yx}(0)$ are ~0.4952 $h/e^2$, ~0.4942 $h/e^2$ and ~0.4899 $h/e^2$ for $m$ =2, 2.5, and 3, respectively. The corresponding $\rho_{xx}(0)$ are ~0.0367 $h/e^2$, ~0.0103 $h/e^2$ and ~0.0240 $h/e^2$ (Figs. S2l to S2n) [23]. Therefore, our magneto-transport measurements suggest a quantum phase transition occurs between the $C = 1$ and $C = 2$ QAH states by varying $m$ from 0 to 3. For devices with $m \geqslant 2$, the remnant $\rho_{xx}(0)$ primarily originates from the MBE growth process. As $m$ increases, additional disorders and/or defects are inevitably introduced into the samples and thus lead to the formation of dissipative channels. This observation aligns with our prior study [22]. We note that kink features are observed in both $\rho_{xx}$ and $\rho_{yx}$ near zero magnetic field and $\mu_0 H$ = ±0.08 T, which are more pronounced for the samples within the quantum phase transition regime, i.e., $1 < m < 2$. During the $\mu_0 H$ sweep process, the eddy current heats the copper sample stage, causing a slight increase in the temperature. Moreover, when $\mu_0 H$ passes through zero magnetic field, the $\mu_0 H$ sweep is paused for 120 seconds at $\mu_0 H$ =0 T. This pause may account for the kink features observed near $\mu_0 H$ =0 T. Note that the kink features near $\mu_0 H$ = ±0.08 T appear only when $\mu_0 H$ sweeps away from $\mu_0 H$ =0 T but disappear when $\mu_0 H$ sweeps towards $\mu_0 H$ =0 T. Because these kink features near $\mu_0 H$ = ±0.08 T are far from the coercive field ($\mu_0 H_c$ ~ ±0.25 T), it is unlikely that these features are a result of the formation of chiral spin textures in the samples.

We attribute the kink features observed near $\mu_0 H = \pm 0.08$ T to a magnetocaloric effect in the samples[33-35]. When $\mu_0 H$ sweeps away from $\mu_0 H = 0$ T, the sample magnetization is realigned, which decreases entropy and releases heat into the electron/phonon system. As a result, the sample temperature increases slightly, leading to the observed kink features in $\rho_{yx}$ near $\mu_0 H = \pm 0.08$ T. We note that this kink is seen in all $C = 2$ QAH samples (i.e., $2 \leq m \leq 3$). This observation suggests that the transport properties of the $C = 2$ QAH samples are more sensitive to thermal heating, presumably due to the enhanced bulk and helical surface carriers introduced by the thicker middle magnetic TI layer. Furthermore, the quantum phase transition between $C = 1$ and $C = 2$ QAH states occurs at zero magnetic field, where the magnetocaloric heating effect is negligible. Therefore, thermal heating is unlikely to induce the observed quantum phase transition between $C = 1$ and $C = 2$ QAH states in our experiment. In addition to the kink features near $\mu_0 H = 0$ T and $\mu_0 H = \pm 0.08$ T, for the samples within the quantum phase transition regime (i.e., $1 < m < 2$), $\rho_{yx}$ exhibits a slight difference between the upward and downward $\mu_0 H$ sweeps near $\mu_0 H_c \sim \pm 0.25$ T. One possible origin of this topological Hall-like hump feature has been attributed to the emergence of chiral spin textures in the samples [26,36].

To confirm the quantum phase transition between the $C = 1$ and $C = 2$ QAH states, we measure the $\mu_0 H$ dependence of $\rho_{xx}$ and $\rho_{yx}$ at different $V_g$s (Figs. S5 to S18) [23]. Figures 2 and S3 show the $V_g$-dependent $\rho_{yx}(0)$ and $\rho_{xx}(0)$. $\rho_{yx}(0)$ is quantized at $\sim h/e^2$ for $0 \leq m \leq 1$ and $\sim h/2e^2$ for $2 \leq m \leq 3$, respectively, over a broad quantized plateau [23]. In both cases, $\rho_{xx}(0)$ vanishes near $V_g^0$ (Fig. S3) [23]. Compared to the electron-doped regime, i.e., $V_g > V_g^0$, the quantization of the QAH sample degrades significantly faster in the hole-doped regime, i.e., $V_g < V_g^0$. This observation can be attributed to the electronic band structure of the magnetic TI, where the magnetic exchange gap is close to the bulk valence band maximum but far from the bulk conduction band minimum, consistent with the prior studies [7,37-39]. Next, we focus on samples within the quantum phase transition regime, i.e., $1 \leq m \leq 2$ (Fig. 2). At $V_g = V_g^0$, the values of $\rho_{yx}(0)$ gradually decrease from $\sim h/e^2$ to $\sim h/2e^2$, with the $\rho_{yx}(0)$ $\sim h/e^2$ plateau narrowing into a distinct peak before returning to the $\rho_{yx}(0)$ $\sim h/2e^2$ plateau. Simultaneously, $\rho_{xx}(0)$ shows a pronounced dip near $V_g^0$ for $m$ =1.1 and 1.2. As $m$ increases, the sharp dip becomes shallower and nearly disappears at $m$ = 1.7. At $V_g = V_g^0$, $\rho_{xx}(0)$ first increases and then decreases at $m$ = 1.7, suggesting the critical point of the quantum phase transition (Fig. 2g). With a further increase in $m$, the dip feature of $\rho_{xx}(0)$ reemerges. This observation suggests the formation of a second chiral edge channel and the emergence of the well-quantized $C$ = 2 QAH state (Fig. 2i). For the $m$ = 1.8 sample, the $\rho_{yx}(0)$ is ~0.51 $h/e^2$ at $V_g = V_g^0$, which is slightly greater than the quantized value. This suggests that the $m$=1.8 sample remains near the quantum phase transition regime, which explains why $\rho_{xx}(0)$ remains large and reaches ~0.31 $h/e^2$ at $V_g = V_g^0$. In addition, the concurrent large $\rho_{xx}(0)$ and nearly quantized $\rho_{yx}(0)$ in magnetically doped TI have been well studied and attributed to the coexistence of quasihelical states and dissipationless chiral edge states in QAH samples [40]. In contrast, the band gap increases for the m = 2 sample, which effectively suppresses the dissipative channels. As a result, the $m$ = 2 sample exhibits a well-quantized $C$=2 QAH state with a significantly smaller $\rho_{xx}(0)$.

Next, we plot the values of $\rho_{yx}(0)$ and $\rho_{xx}(0)$ at $V_g = V_g^0$ as a function of $m$ (Fig. 3a). As $m$ increases from 0 to 3, $\rho_{yx}(0)$ gradually decreases from $\sim h/e^2$ for $0 \leq m \leq 1$ to $\sim h/2e^2$ for $2 \leq m \leq 3$. Meanwhile, $\rho_{xx}(0)$ changes from nearly zero for $0 \leq m \leq 1$ to a finite value for $1 \leq m \leq 2$ and eventually returns to nearly zero for $2 \leq m \leq 3$. We summarize the anomalous Hall (AH) angle $\alpha = \arctan [\rho_{yx}(0)/\rho_{xx}(0)]$, the zero magnetic field Hall conductance $\sigma_{yx}(0)$, and the zero magnetic field longitudinal conductance $\sigma_{xx}(0)$ at $V_g = V_g^0$ as a function of $m$ (Figs. 3b and 3c). During the quantum phase transition between the $C$ =1 and $C$ =2 QAH states, i.e., $1 \leq m \leq 2$, $\rho_{yx}(0)$ is greater than $\rho_{xx}(0)$, and the AH angle α remains consistently larger than 45° (Figs. 3a and 3b). This observation indicates that the chiral edge channel of the original $C$ =1 QAH state persists and coexists with the bulk conducting channels throughout the entire quantum phase transition regime.

To explore the scaling behaviors of the quantum phase transition between the $C$ =1 and $C$ =2 QAH states, we plot the [$\sigma_{xy}(0)$, $\sigma_{xx}(0)$] curves at $V_g = V_g^0$ under different temperatures for all 14 samples with $0 \leq m \leq 3$ (Fig. 3e). The [$\sigma_{xy}(0)$, $\sigma_{xx}(0)$] data points at $T$ =25 mK are highlighted with circles. Note that only the samples with $1 \leq m \leq 2$ are measured over the temperature range from 1K to 25 mK, while the samples with $0 \leq m < 1$ and $2 < m \leq 3$ are measured at $T$ =25 mK. As $T$ decreases, all [$\sigma_{xy}(0)$,$\sigma_{xx}(0)$] curves converge towards two points ($e^2/h$,0) and ($2e^2/h$,0), suggesting that ($e^2/h$,0) and ($2e^2/h$,0) are the two stable fixed points in [$\sigma_{xy}(0)$,$\sigma_{xx}(0)$] space. The [$\sigma_{xy}(0)$, $\sigma_{xx}(0)$] curves recall the RG flow diagram of the plateau phase transition in QH systems (Fig. 3d). In a 2D electron system, the RG flow diagram was introduced to clarify the universality of the QH effect. The characteristic length increases as temperature decreases, and the initial states will be renormalized to different QH states. Each line in Fig. 3d represents the ($\sigma_{xy}$,$\sigma_{xx}$) flow at a fixed magnetic field [13,41,42].

For a specific 2D electron material in real experiments, ($\sigma_{xy}$, $\sigma_{xx}$) flow lines depend on the magnetic field and various material parameters, including carrier density, carrier mobility (i.e., disorder), and scattering mechanism[41,43-45]. For the plateau phase transition between (0, 0) and ($e^2/h$, 0), in disordered GaAs samples with lower carrier mobilities, all RG flow lines are located outside the semicircle centered at ($e^2/2h$, 0) with a radius of ~$e^2/2h$ [44]. However, in GaAs devices with lower disorder and higher carrier mobilities, where the fractional QH states are observed, all RG flow lines lie within the semicircle centered at ($e^2/2h$, 0) with a radius of ~$e^2/2h$ [45]. For our magnetically doped TI penta-layers, Cr dopants introduce substantial disorder and thus significantly reduce the carrier mobility of the samples (<1000 $cm^2$/Vs) [7,11]. Therefore, for all magnetic TI penta-layers with $0 \leqslant m \leqslant 3$, the [$\sigma_{xy}(0)$, $\sigma_{xx}(0)$] flow lines are located outside the semicircle centered at [$\sigma_{xy}(0)$, $\sigma_{xx}(0)$] = ($3e^2/2h$, 0) with a radius of ~$e^2/2h$. Note that each sample transitions from a high-temperature state with large $\sigma_{xx}(0)$ to one of two ground states at ($e^2/h$, 0) and ($2e^2/h$, 0) (Fig. 3e). These flow lines resemble those in disordered GaAs devices[44]. The similarity between our experimental data and the calculated RG flow diagram suggests that the quantum phase transition in a QAH insulator driven by varying $m$ in the middle magnetic TI layer can be described by the same universal scaling behavior with $\mu_0H$-driven plateau transition in a 2D electron gas system. We note that the slight deviation near the middle transition region is likely due to the growth sensitivity of the sample near the critical point (i.e., $m$=1.7).

Next, we will demonstrate the quantum phase transition between the $C$ =1 to $C$ =2 QAH states as a consequence of decoupling the two $C$ =1 QAH states formed in the upper and lower undoped TI layers of the penta-layers. As noted above, the heavily Cr-doped $(Bi,Sb)_2Te_3$ layer acts as a trivial ferromagnetic insulator, supported by the observation of the $C$ =2 QAH state for $m \geq 2$ [22,27]. When we insert a heavily Cr-doped $(Bi,Sb)_2Te_3$ thin layer between the upper and lower TI layers, four gapped Dirac surface states emerge, with two from each undoped TI layer. For $m \leq 1$, the hybridization gap exceeds the magnetic exchange gap for the middle two Dirac surface states, making them topologically trivial and irrelevant to the Chern number $C$. Consequently, only one pair of topological nontrivial interface states is formed, resulting in the $C$ = 1 QAH state in the penta-layers. Note that nontrivial interface state in magnetic TI multilayers refers to the 2D gapped Dirac state that is located at the interface between the undoped TI layer [i.e., $(Bi,Sb)_2Te_3$] and ferromagnetic trivial insulator [i.e., heavily Cr-doped $(Bi,Sb)_2Te_3$ layer] and gives rise to a half-quantized Hall conductance [7,22,27,46]. In contrast to gapless surface/interface states in other nonmagnetic topological materials, which typically rely on symmetry protection, the nontrival interface states in magnetic TI multilayer does not require any symmetry protection [7,22,27,46]. For $m > 1$, the coupling between the upper and lower $C = 1$ QAH layers weakens, reducing the size of the hybridization gap. When the magnetic exchange gap in the two inner surface states surpasses the hybridization gap (i.e., $m \geq 1.7$), these two inner surface states become topologically nontrivial, contributing an additional Hall conductance of $e^2/h$. This results in the realization of the $C = 2$ QAH state in the magnetic TI pentalayers.

The competition between the magnetic exchange and hybridization gaps can be further elucidated by examining the critical temperature and excitation energy gap across all 14 samples. The critical temperature $T_c$ of the QAH state is defined as the temperature at which $\rho_{yx}(0)/\rho_{xx}(0) = 1$ [7,27]. Figures 4a to 4i show the temperature dependence of $\rho_{yx}(0)$ and $\rho_{xx}(0)$ at $V_g = V_g^0$ for the samples with $1 \leq m \leq 2$. For each sample, the $T_c$ value of the QAH state is marked by an arrow. The QAH $T_c$ values are ~6.0 K, ~6.2 K, ~6.1K, ~4.4 K, ~3.8 K, ~2.3 K, ~0.8 K, ~5.6 K, and ~5.8 K for $m$ =1, 1.1, 1.2, 1.3, 1.5, 1.6, 1.7, 1.8, and 2, respectively. Figure 4j shows the QAH $T_c$ values as a function of $m$, with a minimum $T_c$ observed at $m$ =1.7.

Next, we estimate the excitation energy gap $\Delta$ for each sample within the quantum phase transition from its $\sigma_{xx}(0)$-$T$ curve (Fig. S4) [23]. For $T < 300$ mK, the electrical transport behavior is dominated by the variable range hopping due to the localization effect from local disorders [47]. For 300 mK $\leq T \leq$ 1 K, the transport behavior is primarily determined by the effective excitation energy gap $\Delta$, which can be described using the Arrhenius equation: $\sigma_{xx} = \sigma_{xx}^0 e^{-\frac{\Delta}{k_B T}}$. The values of the extracted excitation energy gap Δ are ~49.21 μeV, ~36.53 μeV, ~26.44 μeV, ~17.53 μeV, ~15.78 μeV, ~10.07 μeV, ~7.98 μeV, ~8.98 μeV, and ~18.15 μeV for *m* = 1, 1.1, 1.2, 1.3, 1.5, 1.6, 1.7, 1.8, and 2, respectively. As *m* increases from 1 to 2, the excitation energy gap Δ first decreases, reaching its minimum value at *m*=1.7, and then increases (Fig. 4k). The behaviors of the QAH critical temperature $T_c$ and the excitation energy gap Δ as a function of *m* agree well with our proposed model. Based on this model, the hybridization gap of the middle two Dirac surface states decreases as *m* increases. Finally, the magnetic exchange gap becomes dominant, driving the middle two Dirac surface states into nontrivial interface states, and thus, the entire penta-layer reaches the *C* =2 QAH state.

To further understand the quantum phase transition between the *C* =1 and *C* =2 QAH states, we perform theoretical calculations to simulate the evolution of the energy gap size as a function of *m*. A topological phase transition requires the closing of the band gap, as any two gapped states that can be adiabatically connected without a gap closing must share the same topological property. Therefore, by plotting the band gap in parameter space (*m* and *g* in Fig. S19a), topological phase transition lines can be identified based on where the gap closes [8,48-50]. Moreover, once the Chern numbers for these two phases along a line are identified, the Chern numbers for the corresponding regions in the whole phase diagram (Fig. S19a) are determined. Figure S19a shows the energy gap size of the magnetic TI penta-layer sample as a function of the exchange coupling strength *g* in the magnetic TI layer and its thickness *m* [23]. The dark blue line indicates where the gap closes, separating the *C* = 1 and *C* = 2 QAH states. For *g* =0.18 eV, the band structures of the penta-layers with *m*=1.1, 1.7, and 3 are calculated (Figs. S20a, S20c, and S20e) [23], along with the wavefunction distributions as a function of the vertical position *z* for the lowest four conduction bands at Γ point (Figs. S20b, S20d, and S20f) [23]. The 2$^{nd}$ and 3$^{rd}$ bands are nearly degenerate.

As $m$ increases, a clear band gap closing and reopening feature is observed. Simultaneously, the states located at the interfaces of the middle magnetic TI layer, which corresponds to the wavefunction of the 1$^{st}$ and 4$^{th}$ conduction bands, become less hybridized. We further calculate the Chern number $C$ of the magnetic TI penta-layers as a function of $m$ (Fig. S19b) [23] and confirm that the quantum phase transition between the $C$ =1 to $C$ =2 QAH states occurs at $m$ =1.7, which agrees well with our experimental observations (Figs. 1 to 3).

To summarize, we realize the quantum phase transition between the $C$ = 1 and $C$ = 2 QAH states by varying the middle magnetic TI layer thickness $m$ in magnetic TI penta-layers. The $C$ = 1 QAH state is observed for $m \leq 1$, whereas the $C$ = 2 QAH state emerges for $m \geq 2$. For $1 < m < 2$, the quantum phase transition between the $C$ = 1 and $C$ = 2 QAH states occurs. During the transition, the longitudinal resistance shows a peak near the critical point, and the Hall resistance shows a monotonic decrease from $h/e^2$ to $h/2e^2$. The scaling behavior qualitatively agrees with the RG flows in the QH system, implying a common theory underlying the plateau phase transitions in the QAH and the QH insulators. The quantum phase transition between $C$ = 1 and $C$ = 2 QAH states can be attributed to the weakening of the interlayer coupling between the top and bottom $C$=1 QAH layers. Within the quantum phase transition regime, as $m$ increases, the hybridization gap between the top and bottom $C$=1 QAH layers decreases, eventually giving way to the dominance of the magnetic exchange gap. Our combined experimental measurements and theoretical calculations demonstrate the critical point of the quantum phase transition at $m$ = 1.7. Our work advances our understanding of the quantum phase transition between two QAH states with different Chern numbers and paves the way for developing energy-efficient electronic and spintronic devices based on high Chern number QAH materials.

**Acknowledgments:** This work is primarily supported by the NSF grant (DMR-2241327),

including MBE growth, dilution transport measurements, and theoretical calculations. The device fabrication is supported by the AFOSR grant (FA9550-21-1-0177). The sample characterization is supported by the ARO Award (W911NF2210159). C. Z. C. acknowledges the support from the Gordon and Betty Moore Foundation's EPiQS Initiative (GBMF9063 to C. -Z. C.).

**Data availability:** All data generated in this study is available in Zenodo [51].

**Figures and figure captions:**

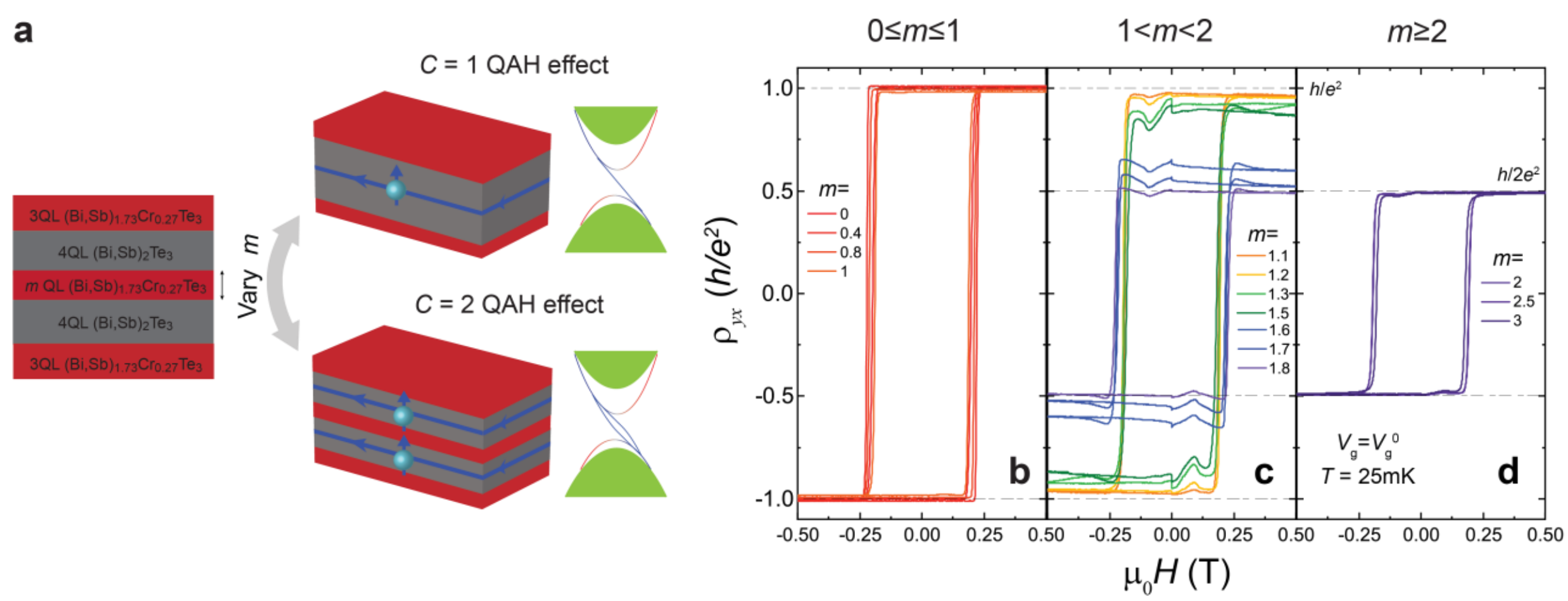

**Fig. 1| Quantum phase transition between *C* = 1 and *C* = 2 QAH states. a**, Left: Schematic of the magnetic TI penta-layer [i.e., 3 QL $(Bi,Sb)_{1.73}Cr_{0.27}Te_3$/4 QL $(Bi,Sb)_2Te_3$/*m* QL $(Bi,Sb)_{1.73}Cr_{0.27}Te_3$/4 QL $(Bi,Sb)_2Te_3$/3 QL $(Bi,Sb)_{1.73}Cr_{0.27}Te_3$]. Right: Schematic of the *m*-change-induced quantum phase transition between $C$ = 1 and $C$ = 2 QAH states. **b-d**, $\mu_0 H$ dependent $\rho_{yx}$ of magnetic TI penta-layers with $0 \leq m \leq 1$(**b**), $1 < m < 2$(**c**), and $2 \leq m \leq 3$(**d**). All measurements are taken at $V_g = V_g^0$ and $T$ =25 mK.

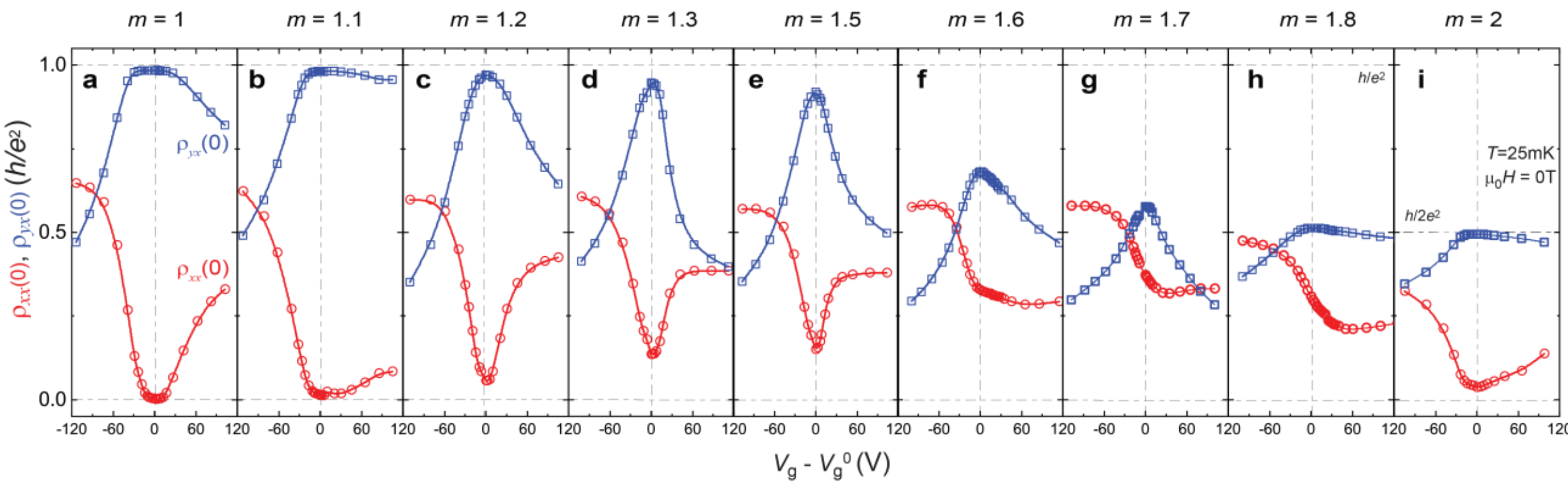

**Fig. 2| Demonstration of the *m*-change-induced quantum phase transition between *C* = 1 and *C* = 2 QAH states. a-i**, ($V_g$–$V_g^0$) dependent $\rho_{yx}(0)$ (blue squares) and $\rho_{xx}(0)$ (red circles) of the magnetic TI penta-layers with *m*=1 (**a**), *m*=1.1 (**b**), *m*=1.2 (**c**), *m*=1.3 (**d**), *m*=1.5 (**e**), *m*=1.6 (**f**), *m*=1.7 (**g**), *m*=1.8 (**h**), and *m*=2 (**i**). All measurements are carried out at *T* =25 mK and $\mu_0 H$ =0 T.

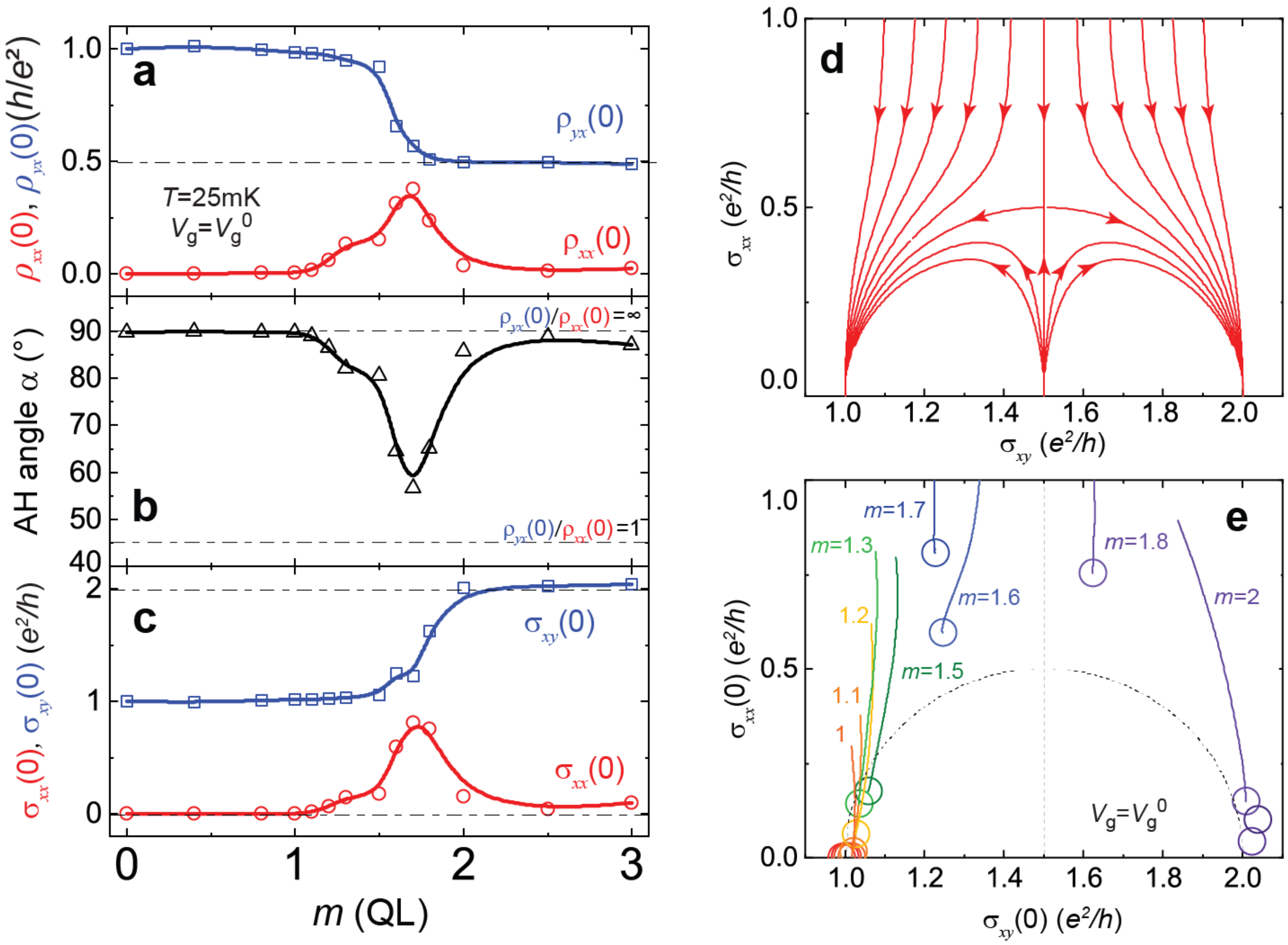

**Fig. 3| Interlayer-coupling-induced quantum phase transition between $C = 1$ and $C = 2$ QAH states. a-c**, $m$ dependence of $\rho_{yx}(0)$ (blue squares) and $\rho_{xx}(0)$ (red circles) (**a**), AH angle α (**b**), and $\sigma_{xy}(0)$ (blue squares) and $\sigma_{xx}(0)$ (red circles) (**c**). All measurements are taken at $V_g=V_g^0$ and $T$=25 mK. **d**, The calculated RG flow diagram of ($\sigma_{xx}$,$\sigma_{xy}$) for a 2D electron system with $1 \leqslant \sigma_{xy} \leqslant 2$. The arrow indicates the flow direction with a lower temperature. **e**, Flow diagram of [$\sigma_{xx}(0)$,$\sigma_{xy}(0)$] for the magnetic TI penta-layers with $0 \leq m \leq 3$. For $1 \leq m \leq 2$, the data are acquired at 25 mK $\leq T \leq$ 1 K. For $0 \leq m < 1$ and $2 < m \leq 3$, the data are acquired at $T$=25 mK. The circle for each sample in (**e**) indicates the data point at $T$=25 mK.

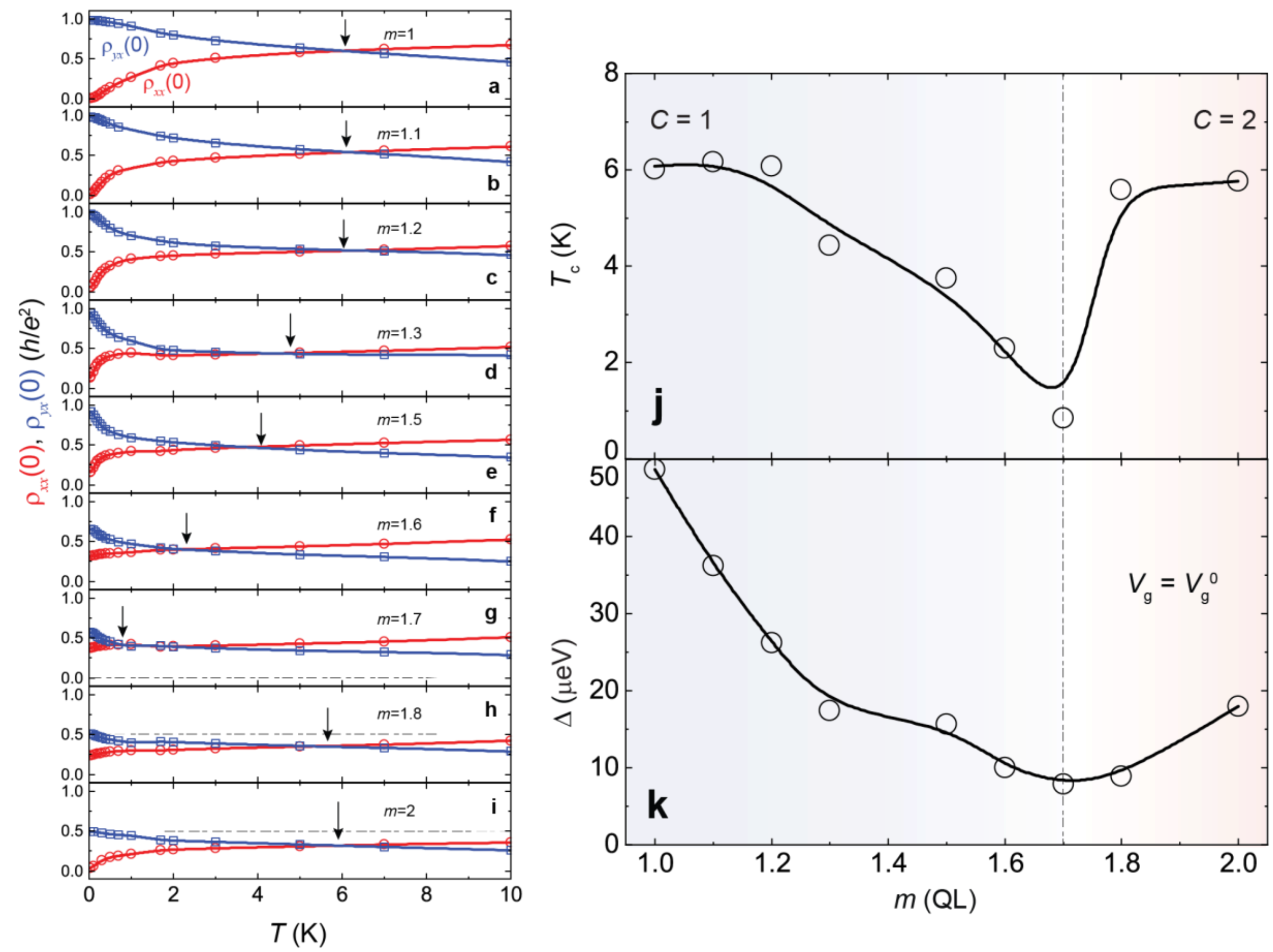

**Fig. 4| Evolution of the QAH critical temperature and excitation gap in magnetic TI penta-layers. a-i**, Temperature dependent $\rho_{yx}(0)$ (blue squares) and $\rho_{xx}(0)$ (red circles) of the magnetic TI penta-layers with $m$ = 1(**a**), $m$ = 1.1(**b**), $m$ = 1.2(**c**), $m$ = 1.3(**d**), $m$ = 1.5(**e**), $m$ = 1.6(**f**), $m$ = 1.7(**g**), $m$ = 1.8(**h**), and $m$ = 2(**i**). All measurements are made at $\mu_0 H$ = 0T. The QAH critical temperature $T_c$ is marked by an arrow for each sample. **j, k,** $m$ dependence of the QAH critical temperature $T_c$ (**j**) and the excitation gap Δ (**k**). The excitation gap Δ is estimated by fitting the data points in the temperature range from 300mK to 1K using the Arrhenius function $\sigma_{xx} = \sigma_{xx}^0 e^{-\frac{\Delta}{k_B T}}$.